\documentclass[11pt]{article}
\usepackage{latexsym}
\begin{document}
\newcommand\lsim{\roughly{<}}
\newcommand\gsim{\roughly{>}}
\newcommand\CL{{\cal L}}
\newcommand\CO{{\cal O}}
\newcommand\half{\frac{1}{2}}
\newcommand\beq{\begin{eqnarray}}
\newcommand\eeq{\end{eqnarray}}
\newcommand\eqn[1]{\label{eq:#1}}
\newcommand\intg{\int\,\sqrt{-g}\,}
\newcommand\eq[1]{eq. (\ref{eq:#1})}
\newcommand\meN[1]{\langle N \vert #1 \vert N \rangle}
\newcommand\meNi[1]{\langle N_i \vert #1 \vert N_i \rangle}
\newcommand\mep[1]{\langle p \vert #1 \vert p \rangle}
\newcommand\men[1]{\langle n \vert #1 \vert n \rangle}
\newcommand\mea[1]{\langle A \vert #1 \vert A \rangle}
\newcommand\bi{\begin{itemize}}
\newcommand\ei{\end{itemize}}
\newcommand\be{\begin{equation}}
\newcommand\ee{\end{equation}}
\newcommand\bea{\begin{eqnarray}}
\newcommand\eea{\end{eqnarray}}
\def\Dsl{\,\raise.15ex \hbox{/}\mkern-12.8mu D}
\newcommand\Tr{{\rm Tr\,}}
\thispagestyle{empty}
\begin{titlepage}
\begin{flushright}
CALT-68-2365\\
\end{flushright}
\vspace{1.0cm}
\begin{center}
{\LARGE \bf  Portfolio Allocation to Corporate Bonds with Correlated Defaults }\\ 
\bigskip
\bigskip\bigskip
{ Mark B. Wise$^a$ and Vineer Bhansali$^b$} \\
~\\
\noindent
{\it\ignorespaces
          (a) California Institute of Technology, Pasadena CA 91125\\

          {\tt wise@theory.caltech.edu}\\
\bigskip   (b) PIMCO, 840 Newport Center Drive, Suite 300\\
               Newport Beach, CA 92660 \\

{\tt   bhansali@pimco.com}
}\bigskip
\end{center}
\vspace{1cm}
\begin{abstract}
This article deals with the problem of optimal allocation of capital to corporate bonds in fixed income portfolios when there is the possibility of correlated defaults. Under fairly general assumptions for the distribution of the total net assets of a set of firms we show that retaining the first few moments of the portfolio default loss distribution gives an extremely good approximation to the full solution of the asset allocation problem. We provide detailed results on the convergence of the moment expansion. We also provide explicit 
results for the inverse problem, i.e. for a given allocation to the set of risky bonds, what is the average risk premium required to make the portfolio optimal. Numerous numerical illustrations exhibit the results for realistic portfolios and utility functions.
\end{abstract}
\vfill
\end{titlepage}

\section{Introduction}
Investors routinely look to the corporate bond market in particular, and spread markets in general, to enhance the performance of their portfolios. However, for every source of excess return over the risk-free rate, there is a source of excess risk. When sources of risk are correlated, the allocation decision to the risky sectors, as well as allocation to particular securities in that sector, can be substantially different from the uncorrelated case. Since the joint probability distribution of returns of a set of defaultable bonds varies with the joint probabilities of default, recovery fraction for each bond, and the number of defaultable bonds, a direct approach to the allocation problem that incorporates all these factors completely can only be attempted numerically within the context of a default model. This approach can have the short-coming of hiding the intuition behind the asset allocation process in practice, which leans very heavily on the quantification of the first few moments, such as the mean, variance and skewness. In this paper, we will take the practical approach of characterizing the portfolio default loss distribution in terms of its moment expansion. Focusing on allocation to corporate bonds (the analysis in this paper can be generalized to any risky sector which has securities with discrete payoffs), we will answer the following questions:
\begin{itemize}
\item In the presence of correlated defaults, how well does retaining the first few moments of the portfolio default loss distribution do, for the portfolio allocation problem, as compared to the more intensive full numerical solution?
\item For different choices of correlations, probabilities of defaults, and number of bonds in the portfolio, how does the optimal allocation to the risky bonds vary?
\item For a target allocation to the risky bond sector, what is the required excess return over the actuarially fair value that is a reasonable compensation for the risk when default correlations are present?
\end{itemize}

In this paper we consider portfolios consisting of risk free assets at a return $r$ and corporate bonds with promised return $c_i$ for firm $i$, and study how correlations between defaults affect the optimal allocation to corporate assets in the portfolio. Let $\hat n_i$ denote a random variable that takes the value $1$ if  company $i$ defaults in the time horizon $\Delta t$ and zero otherwise. $P_i=E[\hat n_i]$ is the probability of firm $i$ defaulting and $d_{ij}=Corr[\hat n_i,\hat n_j]$ is the default correlation between firms $i$ and $j$. 
We will quantify the impact of risk associated with losses from corporate defaults at some single period time horizon $\Delta t$ \footnote{In this paper, our numerical analysis will assume $\Delta t$ to be one year so that default probabilities quoted are annualized ones.} on portfolio allocation. We assume that if company $i$ defaults in the time period $\Delta t$ a fraction $(1-R_i)$ of the portfolio's original investment, in bonds of company $i$ is lost\footnote{ $R_i$ is called the recovery fraction. More explicitly, if the original investment in company $i$ is $1$ dollar then after the time period $\Delta t$ the value of the investment is taken to be $c_i+R_i$ dollars if company $i$ defaults and $1+c_i$ dollars if it survives. Here $c_i$ is the promised corporate return.}. To compensate the investor for default risk, the corporate bond's promised return $c_i$ is greater than that of the risk-free assets in the portfolio. The excess return, or spread of investment in bonds of firm $i$ can be decomposed into two parts. The first part of the spread arises 
as the actuarially fair value of assuming the risk of default. The second part, which we call $\mu_i$, is the excess risk premium that compensates the corporate bond investor above and beyond the probabilistically fair value. In practice, $\mu_i$ can arise due to a number of features not directly related to defaults.  For example, traders will partition $\mu_i$ into two pieces, one for liquidity and one for event risk, $\mu_i = l_i+e_i$.  Given that most low grade bonds may have a full percentage point arising simply from liquidity premia, $\mu_i$ can fluctuate to large values in periods of credit stress. Liquidity $l_i$ is systematic and one would expect it to be roughly equal for a similar class of bonds. Event risk $e_i$ contributes to $\mu_i$ due to the firm's specific vulnerability to factors that affect it (e.g. negative press). In periods of stress $P_i, l_i$ and $e_i$ all increase simulataneously, and the recovery rate expectations $R_i$ fall, leading to a spike in the overall spread.  Since these variables can be highly volatile, the reason behind the portfolio approach to managing credit is to minimize the impact of non-systematic event risk in the portfolio.\footnote{The authors would like to thank David Hinman of PIMCO for enlightening discussions on this topic.}  The risk premium itself is not very stable over time.  Empirical research shows that the excess risk premium might vary from tens of basis points to hundreds of basis points.  For instance, in the BB asset class, if we assume a recovery rate of 50\% and default probability of 2\%, the actuarially fair value of the spread is 100 basis points.  However, it is not uncommon to find actual spreads of the BB class to be 300 bp over treasuries [Altman (1989)].  The excess 200 bp of risk premium can be decomposed in any combination of liquidity premium and event risk premium, and is best left to the judgment of the market.  When the liquidity premium component is small compared to event risk premium, we would expect that portfolio diversification and the methods of this paper are extremely valuable. 

 One other factor needs to be kept in mind when comparing historical spreads to current levels.  In the late eighties and early nineties, the spread was routinely quoted in terms of a treasury benchmark curve.  However, the market itself has developed to a point where spreads are quoted over both the libor swap rate and the treasury rate, and the swap rate has gradually substituted the treasury rate as the risk-free benchmark curve.  This has two impacts.  Firstly, since the swap spread (swap rate minus treasury rate for a given maturity) in US is positive (and of the order of 50 bp as of this writing), the excess spread needs to be computed as a difference to the swap yield curve.  Secondly, the swap-spread itself has been very volatile during the last few years, which leads to an added source of non default related risk in the spread of corporate bonds when computed against the treasury curve.  
Thus, the 200 bp of residual spread is effectively 150 bp over the swap rate in the BB example, of which, for lack of better knowledge, equal amounts may be assumed to arise from liquidity and event risk premium over the long term.  The allocation decision to risky bonds strongly depends on the level of risk-aversion in the investor's utility function, and the required spread for a given allocation will go up nonlinearly as risk aversion increases.

The value of the optimal fraction of the portfolio in corporate bonds, here called $\alpha_{opt}$, cannot be determined without knowing the excess risk premium part of the promised corporate returns. The portfolio average of the excess risk premium, $\mu$, provides the incentive for a risk averse investor to choose corporate securities over risk free assets and so the optimal fraction of corporates in a portfolio $\alpha_{opt}$ is a function of $\mu$ and vanishes at $\mu=0$ (since there is no incentive for a risk averse investor choose the corporate assets when $\mu=0$). Inverting this functional relationship the portfolio average excess risk premium is a function of the optimal fraction of corporate assets, $\mu= \mu (\alpha_{opt})$. 

We explore, using utility functions with constant relative risk aversion, the convergence of the moment expansion for $\alpha_{opt}(\mu)$ and $\mu(\alpha_{opt})$. Our work indicates that, (for $\alpha_{opt}$ less than unity)  $\alpha_{opt}(\mu)$ and $\mu(\alpha_{opt})$ are usually determined by the mean, variance and skewness of the portfolio default loss probability distribution. The sensitivity to higher moments increases as $\alpha_{opt}$ does. Some measures of default risk, for example a VAR analysis\footnote{See for example, Jorion (2001).}, may be more sensitive to the tail of the default loss distribution. We also explore how the optimal portfolio allocation scales with the number of firms. If defaults are random the optimal fraction of corporates increases without bound as the number of firms in the portfolio goes to infinity but when there are default correlations $\alpha_{opt}$ goes to a finite value in that limit.

Historical evidence suggests that on average default correlations increase with the time horizon\footnote{ Zhou (2001) derives an analytic formula for default correlations in a first passage-time default model and finds a similar increase with time horizon.}. For example, Lucas (1995) estimates that over one year, two year and five year time horizons default correlations between $Ba$ rated firms are $2\%$, $6\%$ and $15\%$ respectively. However, the errors in extracting default correlations from historical data are likely to be large since defaults are rare. Also these historical analysis neglect firm specific effects that may be very important for portfolios weighted towards a particular economic sector. Furthermore, in periods of market stress default probabilities and their correlations increase [Das, Freed, Geng and Kapadia (2001)] dramatically.

There are other sources of event risk associated with corporate securities. For example, the markets perception of firm $i$'s probability of default $P_i$ could increase over the time horizon $\Delta t$ resulting in a reduction in the value of its bonds. For a recent discussion on portfolio risk due to downgrade fluctuations see Dynkin, Hyman and Konstantinovsky (2002). Here we do not address the issue of risk associated with fluctuations in the credit spread but rather focus on the risk associated with losses from actual defaults, which for the purposes of this paper, are defined as occurring if the assets of a firm fall below its liabilities.

  In the next section a simple model for default is introduced. The model assumes that the difference between the stochastic assets and liabilities, $\hat a_i$ of firm $i$ is an increasing function of a normal random variable\footnote{Note that $\hat a_i$ itself is not necessarily normal or lognormal.} $\hat z_i$.  Default occurs in the time interval $\Delta t$ if $\hat a_i$ fluctuates below zero. The joint default probabilities are expressed in terms of the correlations and volatilities of the risk variables $\hat z_i$. In section $3$ we set up the portfolio problem. Moments of the fractional corporate default loss probability distribution are expressed in terms of default probabilities and it is shown how these can be used to determine $\alpha_{opt}(\mu)$ and $\mu(\alpha_{opt})$. In section $4$ the impact of correlations on the portfolio allocation problem is studied using sample portfolios where all the firms have the same probabilities of default and the correlations between firms are all the same. Concluding remarks are given in section $5$.  Many of the mathematical details are relegated to an appendix.

\section{A Model For Default}
Assume the $i$th firm in the portfolio has stochastic assets $\hat Y_i$ and debt $\hat F_i$. In the time interval $\Delta t$ a firm defaults if its assets fall below the value of its debt. Thus, the probability of firm $i$ defaulting, in the time interval $\Delta t$, is the probability that $\hat a_i = \hat Y_i-\hat F_i\le 0$. Suppose
\be
\hat a_i=\hat Y_i-\hat F_i=g_i(\hat z_i),
\ee
where $\hat z_i$ is a normal random variable of zero mean and $g_i$ is an increasing function of $\hat z_i$. $\hat z_i$ can be thought of as a ``production" factor variable upon which the net assets of the firm $i$ depend. We impose no further restrictions on the form of the net assets of the firm. For example, for some constants $h_i$ and $q_i$ if $g_i(\hat z_i)=h_i+\hat z_i$, then $\hat a_i$ is normal and if $g_i(\hat z_i)=h_i+q_i\exp(\hat z_i)$ then $\hat a_i$ is lognormal. The statement that default occurs if net assets fall below zero may be restated in terms of the production factor $\hat z_i$ crossing some threshold $T_i$ for firm $i$, i.e.
\be
\hat z_i \le g^{(-1)}(0)=-T_i.
\ee
The probability distribution for the normal variables $\hat z_i$ is completely specified by their variances
\be
\sigma_i^2=E[\hat z_i \hat z_i],
\ee and their correlation matrix
\be
\label{covr}
\xi_{ij}={E[\hat z_i \hat z_j]\over \sqrt{E[\hat z_i \hat z_i]E[\hat z_j \hat z_j]}}.
\ee
Without assuming a particular form for $g_i$ we cannot deduce the volatilities $\sigma_i$ and the correlations $\xi_{ij}$ from the fluctuations of the $\hat a_i$'s. However, we will see later in this section that in this model the default probabilities $P_i$ and their correlations $d_{ij}$ do determine $T_i/\sigma_i$ and $\xi_{ij}$ which therefore have a direct connection to the measures that investors use in quantifying security risk.

This is basically the Merton Model\footnote{No time evolution model for the net assets is specified. Allowing default to occur only at the end of the time interval the $\hat a_i$ can be thought of as the net assets at that time. It is possible to extend the work of this paper to a first passage-time model [Black and Cox (1976), Longstaff and Schwartz (1995), Leland and Toft (1996), {\it etc}.] where default is associated with the first time that $\hat z_i$ crosses a threshold $-T_i$. We will consider the implications of correlated default using a first passage-time model in a further publication.} [Merton (1974)] applied over the time horizon $\Delta t$.
In this model the joint probability that $n$ companies, which we choose to label $1, \ldots  n$, default is given by the following integral of default thresholds and correlations:
\begin{equation}
\label{key}
P_{1 \ldots n} = \frac{1}{(2\pi)^{n \over 2}
\sqrt{\det \xi}}\int_{-\infty}^{-\chi_1} dx_1 \cdots \int_{-\infty}^{-\chi_n} dx_n \exp \left[-\frac{1}{2}\sum_{ij} {x_i \xi^{(-1)}_{ij} x_j}\right],
\end{equation}
where the sum goes over $i,j=1, \ldots ,n$, the scaled default thresholds, or ``equivalent distances to default" are
\be
\chi_i={T_i \over \sigma_i},
\ee
and $\xi^{(-1)}$ is the inverse of the correlation matrix.
For $n$ not too large the integrals in equation (\ref{key}) can be done numerically or, since defaults are rare, analytic results can be obtained using the leading terms in an asymptotic expansion of the integral.

As we mentioned in the introduction, the random variable $\hat n_i$ takes the value $1$ if firm $i$ defaults in the time horizon $\Delta t$ and zero otherwise. The default probabilities in equation (\ref{key}) are expectations of products of these random variables, 
\be
\label{two}
P_{i_1 \ldots i_m}=E[\hat n_{i_1}\cdots \hat n_{i_m}], ~~{\rm when}~~ i_1 \ne i_2 \cdots \ne i_m. 
\ee
$P_i$ is the probability that firm $i$ defaults in the time period $\Delta t$ and $P_{i_1, \ldots ,i_m}$ is the joint probability that the $m$-firms $i_1, \dots i_m$ default in the time period $\Delta t$. 
Since $\hat n_i^2=\hat n_i$ it follows that the correlation of defaults between two different firms (which we choose to label $1$ and $2$) is, 
\be
d_{12}={E[\hat n_1 \hat n_2]-E[\hat n_1]E[ \hat n_2]\over \sqrt{(E[\hat n_1^2]-E[\hat n_1]^2)(E[\hat n_2^2]-E[n_2]^2)}}={ P_{12}-P_1P_2\over \sqrt{P_1(1-P_1)P_2(1-P_2)}}\simeq { P_{12}-P_1P_2\over \sqrt{P_1P_2}},
\ee
When the correlation $\xi_{12}$ and the default probabilities $P_1$ and $P_2$ are small the default correlation $d_{12}$ is also small. There is a simple situation where  default correlations are likely to be large. Suppose company $2$ is dependent on company $1$ so that if company $1$ defaults we know with certainty that company $2$ will default. In that case
$P_{12}=P_1$ and (neglecting terms suppressed by powers of the default probabilities) the default correlation is,
\be
\label{large}
d_{12} \simeq \sqrt{{P_1 \over P_2}}.
\ee
Note that company $2$ can default for reasons unrelated to the health of company $1$ and equation (\ref{large}) still holds.
Suppose company $1$ has default probability $P_1=0.001$ and company $2$ has default probability $P_2=0.01$. Even though both of these are small equation (\ref{large}) implies a large default correlation of $0.33$.

In the next section we consider the problem of portfolio allocation for portfolios consisting of corporate bonds subject to default risk and risk free assets. The implications of default risk are addressed using the model discussed in this section.
Other sources of risk, for example, systematic risk associated with the liquidity part of the excess risk premium, are neglected.

\section{The Portfolio Problem} 
 
The total wealth in a portfolio consisting of risk free assets that return $r$ and corporate assets that are subject to default risk, after time $\Delta t$, is
\be
\label{wealth}
\hat W=W_0\left((1-\alpha)(1+r) +\alpha \sum_i f_i(1+c_i -\hat n_i(1-R_i))\right),
\ee
where, $W_0$ is the initial wealth, $\alpha$ is the fraction of corporate assets in the portfolio, $c_i$ is the return on the i'th corporate asset, $R_i$ is the recovery fraction and $f_i$ denotes the initial fraction of corporate assets in the portfolio that are in firm $i$ ($\sum_i f_i=1$). In equation (\ref{wealth}) the sum over $i$ goes over all $N$ firms in the portfolio. The fractional corporate default loss in the portfolio over the time period $\Delta t$ is the random variable,
\be
\hat l= \sum_i f_i \hat n_i(1-R_i).
\ee
The average fractional default loss is
\be 
E[\hat l]=\sum_i f_i P_i(1-R_i),
\ee
 and the fluctuations of the fractional loss $\hat l$ about its average value is,
\be
\delta \hat l= \hat l-E[\hat l].
\ee
Since the fractional loss $\hat l$ cannot be negative and cannot be greater than one $ -E[\hat l] \le \delta \hat l \le 1-E[\hat l]$. The left hand side of this inequality is typically close to zero and the right hand side is typically close to one since the expected fractional loss is usually small. The mean of $\delta \hat l$ is zero and the probability distribution for $\delta \hat l$ determines the default risk of the portfolio associated with fluctuations of the random variables $ \hat n_i$.  The moments of this probability distribution are, $v^{(m)}=E[(\delta \hat l)^m]$. Using equation (\ref{two}) and the property $\hat n_i^2=\hat n_i$ the  moments $v^{(m)}$ can be expressed in terms of the joint default probabilities. For the first five moments the results are given in the appendix. 

We write the promised corporate return ({\it i.e.}the expected value of the corporate return) for firm $i$ as,
\be
\label{cr}
c_i=r+P_i(1-R_i) +\mu_i ,
\ee
where $\mu_i$ is the excess risk premium  and introduce the portfolio average excess risk premium, 
\be
\mu= \sum_i f_i \mu_i. 
\ee
If an investor holding the portfolio is risk averse, when $\mu =0$ the optimal portfolio has only risk free assets {\it i.e.}, $\alpha_{opt}=0$.

In terms of the quantities we have introduced the total wealth after the time period $\Delta t$ in equation (\ref{wealth}) takes the form,
\begin{equation}
\hat W =W_0\left(1+r+\alpha \mu-\alpha \delta \hat l\right)
\end{equation}

To find out what value of $\alpha$ is optimal a utility function is introduced which characterizes the investor's level of risk aversion. Here we use utility functions of the type $U{_\gamma}(W) =W^{\gamma}/\gamma$ which have constant relative risk aversion\footnote{See, for example, Ingersoll (1987).}, $1-\gamma$. The optimal fraction of corporates, $\alpha_{opt}$ maximizes the expected utility of wealth $E[U{_\gamma}(\hat W)]$. Expanding the utility of wealth in a power series in $\delta \hat l$ and taking the expected value gives
\be
\label{momexp}
E[U{_\gamma}(\hat W)]=(W_0^{\gamma}/\gamma)( 1+r+\alpha \mu)^{\gamma}\left[ 1+\sum_{m=2}^{\infty}{\Gamma (m-\gamma) \over \Gamma (-\gamma)\Gamma(m+1)}\left({\alpha  \over 1+r +\alpha \mu} \right)^m v^{(m)} \right],
\ee
where $\Gamma$ is the Euler Gamma function.
The approximate optimal value of $\alpha$ obtained from truncating the sum in equation (\ref{momexp}) at the $m$'th moment is denoted by $\alpha_m$. The focus of this paper is on portfolios that are not leveraged and have $\alpha_{opt}$ less than unity. We will later see that typically, for such portfolios, the $\alpha_m$ converge very quickly to $\alpha_{opt}$ so that for practical purposes only a few of the moments $v^{(m)}$ need be calculated. 
 
The utility of wealth has explicit dependence on $\mu$ so the optimal value of the fraction of corporates $\alpha_{opt}$ is a function of it {\it i.e.}, $\alpha_{opt}=\alpha_{opt}(\mu)$. Expanding  in a power series about $\mu=0$,
\be
\label{approxalph}
\alpha_{opt} (\mu)= \sum_{n=1}^{\infty}{s_n \over n!} \mu^n.
\ee
The coefficients $s_i$ can be expressed in terms of the moments, $v^{(m)}$. Explicitly, for the first two coefficients\footnote{A derivation of these results for $s_1$ and $s_2$ is given in the appendix.},
\be
\label{ones}
s_1={(1+r) \over (1-\gamma)v^{(2)}},
\ee
and
\be
\label{twos}
s_2=-{(2-\gamma)(1+r)v^{(3)}\over (1-\gamma)^2 (v ^{(2)})^3}.
\ee
The approximation, 
\be
\label{crude}
\alpha_{opt} \simeq s_1 \mu={(1+r)\mu \over (1-\gamma)v^{(2)}} ,
\ee
 has the proper behavior as the various parameters it depends on vary. Increasing the portfolio average excess risk premium $\mu$ increases the fraction of corporates. Decreasing $\gamma$ corresponds to greater risk aversion and gives a lower value of $\alpha_{opt}$. If the excess risk premium $\mu $ vanishes then the optimal portfolio contains only risk free assets. 

Inverting equation (\ref{approxalph}) gives the risk premium as a function of the optimal fraction of corporates, $\mu=\mu(\alpha_{opt})$.  Truncating the sum in equation (\ref{approxalph}) at the second term 
\be
\label{dc}
\mu_{\gamma} (\alpha_{opt}) \simeq {s_1-\sqrt{s_1^2+2 \alpha_{opt} s_2} \over -s_2}={1 \over s_1}\left(\alpha_{opt}-{s_2 \over 2 s_1^2}\alpha_{opt}^2 + \cdots\right).
\ee
We have added a subscript $\gamma$ to emphasize that the excess risk premium depends on the assumed level of risk aversion. Note that equation (\ref{dc}) is not a consequence of the moment expansion but rather relies on an expansion of the utility of wealth in the excess risk premium. If the excess risk premium is large even when the moment expansion for $\alpha_{opt}$ works quite well equation (\ref{dc}) may not be useful.

\section{Sample Portfolios}
Here we consider sample portfolios where the correlations and volatilities of the $\hat r_i$ and the default thresholds are the same for all $N$ firms {\it i.e.,} $\xi_{ij}=\xi$ and $\chi_i=\chi$. Then all the probabilities of default are the same $P_i=P$ and the joint default probabilities are also independent of which firms are being considered,
$P_{i_1 \ldots i_m}=P_{12 \ldots m}$. The corporate returns are also assumed to be independent of firm {\it i.e.,}  $c_i=c$ and $\mu_i=\mu$. We also take the portfolios to contain equal assets in the firms so that, $f_i=1/N$, for all $i$, and assume all the recovery fractions are zero. 

There are two cases where results are easily available without using the moment expansion of the utility function. The first of these is random defaults where the correlations vanish. In this case the probability of a fraction loss of $\hat l=n/N$ is $(1-P)^{N-n}P^n N!/(N-n)!n!$ and so the $m$'th moment of the loss distribution is
\be
v^{(m)}=\sum_{n=0}^N\left({n \over N}-P\right)^m (1-P)^{N-n}P^n {N! \over (N-n)!n!}.
\ee
The expected utility of wealth is
\be
\label{ew}
E[U_{\gamma}(\hat W)]=(W_0^{\gamma}/\gamma)\sum_{n=0}^N (1-P)^{N-n}P^n {N! \over (N-n)!n!}\left(1+r +\alpha \mu -\alpha\left({n \over N}-P \right)\right)^{\gamma}.
\ee
Having the explicit expression for the utility of wealth in equation (\ref{ew}) lets us compare results of the moment expansion for the optimal fraction of corporates $\alpha_m$ with the all orders result, $\alpha_{opt}$. These are presented in Table I in the case where the probability of default is $P=2\%$, the risk free return is $3.5\%$ . According to equation (\ref{cr}) this implies a total promised corporate return of $c= 5.6\%$. Results for different values of the number of firms $N$ and the level of risk aversion $1-\gamma$ are shown in Table I. We also give in columns three and four of Table I the
volatility, ${\rm vol}=\sqrt{v^{(2)}}$, and the skewness, ${\rm skew}=v^{(3)}/ \sqrt{v^{(2)}}^3$, of the portfolio default loss distribution.

Increasing $N$ gives a larger value of the optimal fraction of corporates because diversification reduces risk. This occurs very rapidly with $N$. For $N=1$, $\mu=100 {\rm bp}$ and $\gamma=-4$ the optimal fraction of corporates is only $8.3\%$. By $N = 10$ increasing the number of firms has reduced the portfolio default risk so much that the $100 {\rm bp}$ excess risk premium causes a portfolio that is $83\%$ corporates to be preferred. 

For all the entries in Table I the moment expansion converges very rapidly, although for low $N$ it is the small value of $\alpha_{opt}$ that is driving the convergence. Since in equation (\ref{momexp}) the term proportional to $v^{(m)}$ has a factor of $\alpha^m$ accompanying it we expect good convergence of the moment expansion at small $\alpha$.
\begin{center}
\noindent
Table I: Optimal Fraction of Corporates for Sample Portfolios with Random ($\xi=0$) Defaults and Probability of Default $P=2\%$
\vskip 0.25in
\noindent
$
\bordermatrix{& \gamma & N& {\rm vol}\times 10^3&{\rm skew}&\mu &\alpha_2 & \alpha_3& \alpha_4 & \alpha_5 &\alpha_{opt}\cr
&0.01&1 & 140&6.9&10 {\rm bp}&0.053&0.051&0.051&0.051& 0.051\cr
&0.01&5&63&3.1& 10 {\rm bp}&0.27&0.25&0.25&0.25&0.25\cr
&0.01&10&44& 2.2&10 {\rm bp}&0.53&0.51&0.51&0.51& 0.51\cr
&0.01&15&36&1.8&10 {\rm bp}&0.80&0.77& 0.76&0.76&0.76 \cr
&-4&1 & 140&6.9&100 {\rm bp}&0.11&0.086& 0.083&0.083& 0.083  \cr 
&-4&5 &63&3.1&100 {\rm bp}&0.54&0.43&0.42&0.41&0.41\cr 
&-4&10& 44&2.2&100 {\rm bp}&1.1&0.88&0.84&0.83&0.83\cr
}$
\end{center}
\vskip0.25in

The focus of this paper is on portfolios that are not leveraged and have $\alpha_{opt}<1$. But a value $\alpha_{opt} > 1$ is not forbidden when finding the maximum of the expected utility of wealth. This occurs at lowest order in the moment expansion in the last row of Table I.

The other situation where the optimal portfolio can be found without using the moment expansion is when the correlations are maximal, {\it i.e.} $\xi=1$. Then all the companies must default together and so for any N the utility of wealth is given by equation (\ref{ew}) with $N=1$. Explicitly, for maximal correlations,
\begin{eqnarray}
E[U_{\gamma}(\hat W)]&=&(W_0^{\gamma}/\gamma)\left[(1-P)\left(1+r +\alpha \mu +\alpha P\right)^{\gamma} \right.\nonumber \\
&+&\left. P\left(1+r +\alpha \mu -\alpha (1-P)\right)^{\gamma}\right].
\end{eqnarray}
Correlations make diversification less effective and in the extreme case of maximal correlations the optimal fraction of corporates is independent of the number of firms.

Including correlations makes the moment expansion converge slower. The situation will never be worse than the maximal
correlation case and so it is worth examining its convergence. This is done in Table II. Here we take, $\xi=1$, 
$r=0.035$ and $P=2.0\%$. Results are presented for $\gamma=0.01$ and $\gamma=-4$ and excess risk premiums 
between $10 {\rm bp}$ and $400 {\rm bp}$.

With $\gamma=0.01$ the excess risk premium $\mu(54\%)$ is $200 {\rm bp}$ and for that case $\alpha_5$ is within $4\%$ of the correct value. Just including the variance $v^{(2)}$ gives you a result that is off by almost a factor of two but $\alpha_3$ is within $30\%$ of the correct value. In realistic cases where correlations are significantly less than unity the convergence of the moment expansion should be substantially better. 

Note that for $\gamma=0.01$ and excess risk premium of $400{\rm bp}$, gives an optimal portfolio with
almost $75\%$ corporates. This excess risk premium is too small for most investors to risk, with a $2\%$ probability, losing $75\%$ of their wealth. With $\gamma=-4$ the investor requires an excess risk premium of $46.6 \%$ before the optimal fraction of corporates reaches $75\%$.
\vskip0.25in
\begin{center}
Table II: Worst Case Convergence of Moment Expansion, $\xi=1$, for Probability of Default $P=2\%$.
\vskip 0.25in
\noindent
$
\bordermatrix{& \gamma &{\rm vol}\times 10^3&{\rm skew}& \mu&\alpha_2 & \alpha_3& \alpha_4 & \alpha_5 &\alpha_{opt} \cr
&0.01&140&6.9&10 {\rm bp}&  0.053 & 0.051 & 0.051 & 0.051 & 0.051 \cr
 &0.01&140&6.9&100 {\rm bp}&  0.54 & 0.39 & 0.37 & 0.36 & 0.36\cr 
 &0.01&140&6.9&200 {\rm bp}&  1.1 & 0.68 & 0.59 & 0.56 &0.54 \cr
 &0.01&140&6.9&300 {\rm bp}&  1.8 & 0.92 & 0.77 & 0.70 &0.66\cr
 &0.01&140&6.9&400 {\rm bp}&  2.6 & 1.2 & 0.92 & 0.83 &0.74 \cr
 &-4&140&6.9&100 {\rm bp}& 0.11 & 0.086 & 0.083 & 0.083 & 0.083\cr 
& -4&140&6.9 &200{\rm bp}& 0.21 & 0.15 & 0.14 & 0.14 &0.14\cr
& -4&140&6.9&300{\rm bp}&  0.33 & 0.21 & 0.19 & 0.18 &0.18\cr
& -4&140&6.9&400{\rm bp}&  0.45 & 0.26 & 0.23 & 0.22 &0.21\cr
}$
\end{center}
\vskip0.25in

Finally we consider the more typical case $\xi \ne 0,1$. The formulae used to determine the joint default probabilities and the moments $v^{(2)}-v^{(5)}$ are given in the appendix and using them we calculate $\alpha_2-\alpha_5$. 
\newpage
\vskip0.25in
\begin{center}
Table IIIa:  Moment Expansion for Optimal Portfolio with Correlated Defaults. The Probability of Default is $P=2\%$.
\vskip 0.25in
$
\bordermatrix{&\gamma&N&\xi&d_{ij}&{\rm vol}\times 10^3&{\rm skew}&\mu&\alpha_2&\alpha_3&\alpha_4&\alpha_5\cr
&0.01&10&0.50&0.152&68&5.1&10{\rm bp}&0.22&0.21&0.21&0.21\cr
&0.01&10&0.45& 0.126&64&4.8&10{\rm bp}&0.25&0.23&0.23&0.23\cr
&0.01&10& 0.40&0.103&61&4.5&10{\rm bp}&0.28&0.26&0.26&0.26 \cr
&0.01&10&0.35 &0.0823&58&4.2&10{\rm bp}&0.31&0.29&0.29& 0.29 \cr
&0.01&10&0.30&0.0645&56&3.8&10{\rm bp}&0.34&0.32&0.32&0.32\cr
&0.01&10&0.25&0.0491&53&3.5&10{\rm bp}&0.37 & 0.35 & 0.35 & 0.35 &\cr
&0.01&50&0.50&0.152&58&5.5&10{\rm bp}&0.32 & 0.29 & 0.29 & 0.29\cr
&0.01&50&0.45& 0.126&53&5.2 &10{\rm bp}&0.37 & 0.34 & 0.34& 0.34\cr
&0.01&50& 0.40&0.103&49&4.8&10{\rm bp}&0.44 & 0.41& 0.40 & 0.40 \cr
&0.01&50&0.35 &0.0823&44&4.4&10{\rm bp}&0.53 & 0.49 & 0.48& 0.48 \cr
&0.01&50&0.30&0.0645&40&4.0&10{\rm bp}&0.64 & 0.59 & 0.58 & 0.58 \cr
&0.01&50&0.25&0.0491&37&3.5&10{\rm bp}&0.37 &0.78 & 0.72 & 0.71\cr
&0.01&100&0.50&0.152&56&5.6&10{\rm bp}&0.33 & 0.30 & 0.30& 0.30\cr
&0.01&100&0.45& 0.126&51&5.3&10{\rm bp}& 0.40 & 0.36 & 0.36& 0.36\cr
&0.01&100& 0.40&0.103&47&4.9&10{\rm bp}& 0.48 & 0.44& 0.43 & 0.43\cr
&0.01&100&0.35 &0.0823&42&4.6&10{\rm bp}&0.58 & 0.53 & 0.52& 0.52\cr
&0.01&100&0.30&0.0645&38&4.1&10{\rm bp}& 0.72 & 0.66 & 0.65 & 0.65 \cr
&0.01&100&0.25&0.0491&34&3.6&10{\rm bp}& 0.91& 0.83 & 0.82 & 0.82 \cr
}$
\end{center}
\begin{center}
Table IIIb:  Moment Expansion for Optimal Portfolio with Correlated Defaults. The Probability of Default is $P=2\%$.
\vskip 0.25in
$
\bordermatrix{&\gamma &N&\xi&d_{ij}&{\rm vol}\times 10^3&{\rm skew}&\mu&\alpha_2&\alpha_3&\alpha_4&\alpha_5\cr
&-4&10&0.50&0.152&68&5.1&100{\rm bp}& 0.45 & 0.34 & 0.31 & 0.31 \cr
&-4&10&0.45& 0.126&64&4.8&100{\rm bp}&0.50 & 0.38 & 0.35& 0.34\cr
&-4&10& 0.40&0.103&61&4.5&100{\rm bp}&0.56 & 0.42& 0.39 & 0.38 \cr
&-4&10&0.35 &0.0823&58&4.2&100{\rm bp}&0.62 & 0.47 & 0.44& 0.43 \cr
&-4&10&0.30&0.0645&56&3.8&100{\rm bp}&0.69 & 0.52 & 0.48 & 0.47 \cr
&-4&10&0.25&0.0491&53&3.5&100{\rm bp}&0.75 & 0.57 & 0.54 & 0.53  &\cr
&-4&50&0.50&0.152&58&5.5&100{\rm bp} &0.64 & 0.45 & 0.41 & 0.40 \cr
&-4&50&0.45& 0.126&53&5.2&100{\rm bp} &0.76 & 0.53 & 0.48& 0.47\cr
&-4&50& 0.40&0.103&49&4.8&100{\rm bp}&0.91 & 0.63& 0.57 & 0.55 \cr
&-4&50&0.35 &0.0823&44&4.4&100{\rm bp}&1.1 & 0.76 & 0.68& 0.66 \cr
&-4&50&0.30&0.0645&40&4.0&100{\rm bp}&1.3 & 0.92 & 0.83 & 0.79 \cr
&-4&50&0.25&0.0491&37&3.5&100{\rm bp}&1.7 & 1.1 & 1.0 & 0.98 \cr
&-4&100&0.50&0.152&56&5.6&100{\rm bp}&0.67 & 0.47 & 0.43& 0.41\cr
&-4&100&0.45& 0.126&51&5.3&100{\rm bp}& 0.81 & 0.56 & 0.51& 0.49\cr
&-4&100& 0.40&0.103&47&4.9&100{\rm bp}& 0.98 & 0.67& 0.60 & 0.68 \cr
&-4&100&0.35 &0.0823&42&4.6&100{\rm bp} &1.2 & 0.99 & 0.95& 0.94 \cr
&-4&100&0.30&0.0645&38&4.1&100{\rm bp}&1.5 & 1.0 & 0.91 & 0.86  \cr
&-4&100&0.25&0.0491&34&3.6&100{\rm bp}& 1.9 & 1.3 & 1.1 & 1.1 \cr
}$
\end{center}

 Tables III give the results of this calculation for portfolios with $r=3.5\%$ and $P=2\%$. The levels of risk aversion used are $\gamma=0.01$ and $\gamma=-4$ and respective excess risk premiums are taken to be $\mu=10{\rm bp}$ and $\mu=100{\rm bp}$. The number of firms in the portfolios are taken to be $N=10$, $50$ and $100$. Unlike Table I the defaults are not random and values of $\alpha_2-\alpha_5$ are given for correlations $\xi$ ranging from $0.50$ to $0.25$. The default correlations $d_{ij}$ corresponding to these choices of $\xi$ are listed in the fourth column of Tables III. The volatility and skewness of the portfolio default loss distribution are given in columns five and six. The joint default probabilities used to construct Tables III are given in the appendix.

For all the entries in Tables III the convergence of the moment expansion is good. In most cases the second moment gives a reasonable approximation and by the third moment it is usually quite accurate. The convergence is better at lower risk aversion and lower values of $\alpha_{opt}$. 

Correlations dramatically effect the dependence of the optimal portfolio allocation on the total number of firms. For the $\xi=0.5$ entries in Table IIIa the optimal fraction of corporates is, $0.21$, $0.29$, and $0.30$ for $N=10$, $50$ and $100$ respectively. Increasing the number of firms beyond $50$ only results in a small increase in the optimal fraction of corporates. For random defaults ({\it i.e.}, $\xi=0$) the optimal fraction of corporates goes to infinity as the number of firms in the portfolio goes to infinity. That is because the moments of the portfolio loss distribution go to zero as $N \rightarrow \infty$. For example, with $\xi=0$ the variance of the default loss distribution is,
\be
v^{(2)}={P(1-P) \over N},
\ee
and the skewness of the default loss distribution is
\be
{v^{(3)} \over (v^{(2)})^{3/2}}={1 \over \sqrt{NP(1-P)}}(1-2P).
\ee
For random defaults as $N \rightarrow \infty$ the distribution for $\delta \hat l$ approaches the trivial one where $\delta \hat l=0$ occurs with unit probability.
However for $\xi >0$ the moments $v^{(m)}$ go to non-zero values in the limit $N \rightarrow \infty$ and the default loss distribution remains non-trivial and non-normal.

The examples in this section have all used a default probability of $2\%$, however, the convergence of the moment expansion is similar for significantly larger default probabilities. Using $r=3.5\%$, $P=10\%$, $\gamma=-4$, $N=100$ $\mu= 300 {\rm bp}$ and $\xi=0.25$ we find that: $\alpha_2=0.66$, $\alpha_3=0.51$, $\alpha_4=0.48$ and $\alpha_5=0.47$ Again the convergence of the moment expansion is quite good. 

The impact of changing the probability of default on the portfolio average excess risk premium needed to have an optimal portfolio with $25\%$ corporates is studied in Table IV. It contains approximate values of $\mu (25\%)$ for default probabilities $P=2\%$, $4\%$, $6\%$, $8\%$ and $10\%$. Other parameters are: $r=3.5\%$, $\xi=0.25$ and $\gamma=-4$. In Table IV $\mu ^{(1)} (25\%)$ is the excess risk premium that follows from just keeping the variance ({\it i.e.} only $s_1$ in equation (\ref{dc})) and $\mu ^{(2)} (25\%)$ is the value that follows from keeping both the variance and skewness ({\it i.e.} both $s_1$ and $s_2$ in equation (\ref{dc})). Note that even though we are holding the `production" factor variable correlations $\xi$ fixed in Table IV the default correlations change because we are changing the probabilities of default.

Correlations change the behavior of $\mu(\alpha_{opt})$ as the number of firms in the portfolio gets large. For uncorrelated defaults the portfolio average excess risk premium goes to zero as $N \rightarrow \infty$ but with correlations it goes to a nonzero value in this limit. For the $P=10\%$ entries in Table IV the required excess risk premium to achieve an optimal portfolio with $25\%$ corporates drops from $264{\rm bp}$ to $157 {\rm bp}$ to $145{\rm bp}$ in going from $10$ to $50$ to $100$ firms. For $N=10,000$ it is $133{\rm bp}$ indicating that going beyond $100$ firms can further reduce $\mu(25\%)$ by only about $10\%$.
\vskip0.25in
\begin{center}
Table IV:  Approximate Values for $\mu(25\%)$ for Different Default Probabilities. 
\vskip 0.25in
$
\bordermatrix{&\gamma&N&P&\xi&d_{ij}&{\rm vol}\times 10^3& {\rm skew}&\mu^{(1)}(25\%)  &\mu^{(2)}(25\%)\cr
 &-4&10&2\%&0.25&0.0491&53&3.5 & 34{\rm bp}  &41{\rm bp} \cr
 &-4&10&4\%&0.25&0.0691&79&2.6 & 75 {\rm bp}&92{\rm bp}\cr
 &-4&10&6\%&0.25&0.0833&99& 2.2& 120{\rm bp}& 149{\rm bp}\cr
 &-4&10&8\%&0.25&0.0945&117& 1.9& 164 {\rm bp}&207 {\rm bp}\cr
 &-4&10&10\%&0.25&0.104&132& 1.7& 210{\rm bp} & 264{\rm bp}\cr
 &-4&50&2\%&0.25&0.0491&37&3.5 & 16{\rm bp}  &18{\rm bp}\cr
 &-4&50&4\%&0.25&0.0691&58& 2.7& 41 {\rm bp}&47{\rm bp}\cr
 &-4&50&6\%&0.25&0.0833&76&2.3 &69{\rm bp}& 81{\rm bp}\cr
 &-4&50&8\%&0.25&0.945&91&2.0 & 100{\rm bp}&118 {\rm bp}\cr
 &-4&50&10\%&0.25&0.104&105&1.8 & 132{\rm bp} &157 {\rm bp}\cr
&-4 &100&2\%&0.25&0.0491&34&3.6 & 15{\rm bp}  &5.9{\rm bp} \cr
&-4 &100&4\%&0.25&0.0691&55&2.8 & 36 {\rm bp}&42{\rm bp}\cr
&-4 &100&6\%&0.25&0.0833&72& 2.3& 63{\rm bp}& 73{\rm bp}\cr
&-4 &100&8\%&0.25&0.0945&87&2.0 & 92 {\rm bp}&108 {\rm bp}\cr
&-4 &100&10\%&0.25&0.104&101&1.8& 122{\rm bp} & 145{\rm bp}\cr
}$
\end{center}
The approximate values of $\mu(\alpha_{opt})$ in Table IV are close to the correct values. For example, taking $P=10\%$ and $N=100$ we find that an excess risk premium $\mu=145 {\rm bp}$ gives: $\alpha_2=0.30$, $\alpha_3=0.26$, $\alpha_4=0.26$ and $\alpha_5=0.26$.

For $\gamma=-4$, $\alpha_{opt} >25\%$ and $\xi=0.25$, an expansion in the risk premium does not converge fast enough for equation (\ref{dc}) to be useful for $\mu(\alpha_{opt})$. However, one can still find $\mu(\alpha_{opt})$ using the moment expansion for $\alpha_{opt}$. For example with $\gamma=-4$, $N=100$, $\xi=0.25$, $r=3.5\%$ and $P=10\%$ we find that $\mu(50\%)=330{\rm bp}$.

\section{Concluding Remarks}

Default correlations have an important impact on portfolio default risk. Given a default model the tail of the default loss probability distribution is difficult to compute, often involving numerical simulation\footnote{See, for example, Duffie and Singleton (1999) and Das, Fong and Geng (2001).} of rare events. The first few moments of the default loss distribution are easier to calculate, and in the default model we used this involved some simple numerical integration. More significantly, the first few terms in the moment expansion have a familiar meaning. Investors are used to working with the classic mean, variance and skewness measures and have developed intuition for them and confidence in them. In this paper we studied the utility of the first few moments of the default loss probability distribution for the portfolio allocation problem.

The default model we use assumes that the assets minus liabilities of firm $i$, $\hat a_i$, is a function of a normal random variable. However, the $\hat a_i$ themselves are not necessarily normal or lognormal and their probability distribution can have fat tails. When the $\hat a_i$ are correlated the portfolio default loss distribution does not approach a trivial\footnote{The trivial distribution has $\delta \hat l=0$ occuring with unit probability.} or normal probability distribution as the number of firms in the portfolio goes to infinity. Correlations dramatically decrease the effectiveness of increasing the number of firms, $N$, in the portfolio to reduce portfolio default risk. 

The value of the optimal fraction of corporate assets, $\alpha_{opt}$, cannot be determined without knowing the portfolio average of the excess risk premium part of the corporate returns, $\mu$. It provides the incentive for a risk averse investor to choose corporate assets over risk free assets and so $\alpha_{opt}$ is a function of $\mu$ which vanishes at $\mu=0$. Inverting this functional relationship the portfolio average excess risk premium is a function of the optimal fraction of corporate assets, $\mu = \mu (\alpha_{opt})$. This is the amount of ``free money" over the actuarially fair value of the investment an investor that desires an optimal portfolio requires before taking on the added risk of having a fraction $\alpha_{opt}$ of corporate securities in the portfolio. We explored the convergence of the moment expansion for $\alpha_{opt}(\mu)$ and $\mu(\alpha_{opt})$.  Our work indicates that, for $\alpha_{opt}$ less than unity, the convergence of the moment expansion is quite good. The convergence of the moment expansion gets poorer as $\alpha_{opt}$ gets larger and as the level of risk aversion gets larger. 

The values of $\alpha_{opt}(\mu)$ and $\mu (\alpha_{opt})$ depend on the utility function used. In this paper we used utility functions with constant relative risk aversion, $1-\gamma$. It is possible to make other choices and while the quantitative results will be different for most practical purposes we expect that the general qualitative results should continue to hold. 

\vskip0.25in

\noindent
{\Large{\bf References}}
\vskip0.25in

\noindent
Altman, E.I., (1989) {\it Measuring Corporate Bond Mortality and Performance}, Journal of Finance, Vol. 44, No. 4, 909-921.

\vspace{0.2cm}

\noindent
Black, F., and Cox, J. (1976) {\it Valuing Corporate Securities: Some Effects of Bond Indenture Provisions}, Journal of Finance, 31, 351-367.

\vspace{0.2cm}

\noindent
Das, S., Fong, G. and Geng, G. (2001) {\it The Impact of Correlated Default Risk on Credit Portfolios}, working paper, Department of Finance Santa Clara University and Gifford Fong Associates.

\vspace{0.2cm}

\noindent
Das, S., Freed, L., Geng, G. and Kapadia, N. (2001) {\it Correlated Default Risk}, working paper, Department of Finance Santa Clara University and Gifford Fong Associates.

\vspace{0.2cm}

\noindent
Duffie, D. and Singleton, K. (1999) {\it Simulating Correlated Defaults}, working paper, Stanford University Graduate School of Business.

\vspace{0.2cm}

\noindent
Dynkin, L., Hyman, J. and Konstantinovsky, V. (2002) {\it Sufficient Diversification in Credit Portfolios}, Lehman Brothers Fixed Income Research.

\vspace{0.2cm}

\noindent
Ingersoll, J. (1987) {\it Theory of Financial Decision Making}, Rowman and Littlefield Publishers Inc.

\vspace{0.2cm}

\noindent
Jorion, P. (2001) {\it Value at Risk}, Second Edition, McGraw-Hill Inc.

\vspace{0.2cm}

\noindent
Leland, H. and Toft, K. (1996) {\it Optimal Capital Structure, Endogenous Bankruptcy and the Term Structure of Credit Spreads}, Journal of Finance, 51, 987-1019.

\vspace{0.2cm}

\noindent
Longstaff, F. and Schwartz, E. (1995) {\it A Simple Approach to Valuing Risky Floating Rate Debt}, Journal of Finance, 50, 789-819.

\vspace{0.2cm}
\noindent
Lucas, D. (1995), {\it Default Correlation and Credit Analysis}, Journal of Fixed Income, March, 76-87.

\vspace{0.2cm}

\noindent
Merton, R. (1974), {\it On Pricing of Corporate Debt: The Risk Structure of Interest Rates}, Journal of Finance 29, 449-470.

\vspace{0.2cm}

\noindent
Rebonato, R. and  J\"{a}ckel, P. (2000) {\it The Most General Methodology for Creating a Valid Correlation Matrix for Risk Management and Option Pricing Purposes}, The Journal of Risk, Vol. 2, No. 2, 17-26.

\vspace{0.2cm}

\noindent
Zhou, C. (2001) {\it An Analysis of Default Correlations and Multiple Defaults}, The Review of Financial Studies, Vol. 12, No. 2, 555-576.

\vspace{0.2cm}

\appendix

\section{Selected Mathematical Details}

The moments of probability distribution for fluctuations in the fractional default loss are defined by,
\be
v^{(m)}=E[(\delta \hat l)^m].
\ee
It is convenient to introduce the quantity,
\be
\tilde f_i=f_i(1-R_i).
\ee
Explicit expressions for the first five moments in terms of the default probabilities, $P_{i_1,\ldots i_m}$ are,
\be
\label{mom2}
v^{(2)}=\sum_{i \ne j}\tilde f_i \tilde f_j P_{ij}+\sum_i \tilde f_i^2 P_i -E[\hat l]^2, 
\ee 

\be
\label{mom3}
v^{(3)}=\sum_{i \ne j \ne k} \tilde f_i \tilde f_j \tilde f_k P_{ijk} +3\sum_{i \ne j} \tilde f_i^2 \tilde f_j P_{i j} +\sum_i \tilde f_i^3P_i-3 v^{(2)}E[\hat l]-E[\hat l]^3,
\ee

\begin{eqnarray}
\label{mom4}
v^{(4)} &=&\sum_{i \ne j \ne k \ne l} \tilde f_i \tilde f_j \tilde f_k \tilde f_l P_{ijkl} +6 \sum_{i \ne j \ne k }\tilde f_i^2 \tilde f_j \tilde f_k P_{ijk}+4 \sum_{i \ne j} \tilde f_i^3 \tilde f_j P_{ij} \nonumber \\
&+&3 \sum_{i\ne j}\tilde f_i^2 \tilde f_j^2 P_{ij} +\sum_i \tilde f_i^4 P_i -4v^{(3)} E[\hat l]-6 v^{(2)}E[\hat l]^2-E[\hat l]^4,
\end{eqnarray}
and
\begin{eqnarray}
\label{mom5}
v^{(5)} &=&\sum_{i \ne j \ne k \ne l \ne m} \tilde f_i \tilde f_j \tilde f_k \tilde f_l \tilde f_m P_{ijklm}+10 \sum_{i \ne j \ne k \ne l} \tilde f_i^2 \tilde f_j \tilde f_k \tilde f_l P_{ijkl}+10 \sum_{i \ne j \ne k }\tilde f_i^3 \tilde f_j \tilde f_k P_{ijk} \nonumber \\
&+&15 \sum_{i \ne \ne j \ne k}\tilde f_i \tilde f_j^2 \tilde f_k^2 P_{ijk}+10 \sum_{i \ne j}\tilde f_i^3 \tilde f_j^2 P_{ij}+5 \sum_{i \ne j} \tilde f_i^4 \tilde f_j P_{ij}+\sum_i \tilde f_i^5 P_i  \nonumber \\
&-& 5 v^{(4)} E[ \hat l]-10 v^{(3)} E[ \hat l]^2-10 v^{(2)}E[ \hat l]^3-E[\hat l]^5.
\end{eqnarray}
Even if the defaults are random the probability distribution for $\delta \hat l$ has non zero higher moments. As the number of firms in the portfolio increases the importance of this goes down in comparison with the effects of correlations. 

The expression for the variance, $v^{(2)}$, given there can be rewritten as,
\be
\label{momhaz}
v^{(2)}=\sum_{i \ne j}\tilde f_i\tilde f_j(P_{ij}-P_iP_j)+\sum_i \tilde f_i^2P_i(1-P_i).
\ee
Equation (\ref{momhaz}) can be expressed in terms of the default probabilities $P_i$ and default correlations $d_{ij}$. Assuming that the $P_i$ are small,
\be
v^{(2)}\simeq  \left(\sum_{i \ne j}\tilde f_i \tilde f_j \sqrt{ P_i P_j}d_{ij}+\sum_i \tilde f_i^2P_i \right).
\ee

The maximum of the utility of wealth occurs at a value of $\alpha $ where its first derivative vanishes. Differentiating the utility of wealth with respect to $\alpha$ and equating it to zero gives,
\be
\label{deriv}
0={\mu \over 1+r}-{(1-\gamma) v^{(2)} \alpha_{opt} \over (1+r)^2}-{(2-\gamma)(1-\gamma)v^{(3)} \alpha_{opt}^2 \over 2(1+r)^3}+  \ldots .
\ee
Using the expansion of $\alpha_{opt}$ is powers of the excess risk premium given in equation (\ref{approxalph}) this becomes,
\begin{eqnarray}
\label{next}
0&=&{\mu \over 1+r}-{(1-\gamma)  v^{(2)} s_1 \mu \over (1+r)^2}-{(1-\gamma) v^{(2)} s_2 \mu^2 \over 2(1+r)^2} \nonumber \\
&-&{(2-\gamma)(1-\gamma)v^{(3)} s_1^2 \mu^2 \over 2(1+r)^3}+\ldots,
\end{eqnarray}
where the ellipses denote terms higher order\footnote{In equation (\ref{deriv}) the ellipses denote terms that give rise to effects higher than order $\mu^2$ in equation (\ref{next}).}in the excess risk premium $\mu$. To derive equations (\ref{ones}) and (\ref{twos}) we note that since $\mu$ is arbitrary the coefficients of $\mu$ and $\mu^2$ must vanish yielding,
\be
s_1={1+r \over (1-\gamma) v^{(2)}},
\ee
and
\be
s_2=-{(2-\gamma)s_1^2 v^{(3)} \over v^{(2)} (1+r)}=-{(2-\gamma)(1+r)v^{(3)}\over (1-\gamma)^2 (v ^{(2)})^3}.
\ee

For sample portfolios discussed in section 4  all the firms have the same probability of default and $f_i=\tilde f_i=1/N$. In that case, to express the moments $v^{(m)}$ in terms of the default probabilities we use equations (\ref{mom2}) to (\ref{mom5}) which yield the following results,
\be
v^{(2)}={(N-1) \over N} P_{12}+{1 \over N}P-P^2,
\ee
\be
v^{(3)}={(N-1)(N-2) \over N^2}P_{123}+3{(N-1) \over N^2} P_{12} +{1 \over N^2}P-3v^{(2)}P -P^3,
\ee
\begin{eqnarray}
v^{(4)}&=&{(N-1)(N-2)(N-3) \over N^3}P_{1234}+6{(N-1)(N-2) \over N^3}P_{123}+7{(N-1) \over N^3} P_{12} \nonumber \\
&+& {1 \over N^3}P-4v^{(3)}P-6v^{(2)}P^2-P^4,
\end{eqnarray}
and
\begin{eqnarray}
v^{(5)}&=&{(N-1)(N-2)(N-3)(N-4)\over N^4}P_{12345}+10{(N-1)(N-2)(N-3) \over N^4}P_{1234} \nonumber \\
&+&35{(N-1)(N-2) \over N^4}P_{123}+15{(N-1)P_{12} \over N^4}+{1 \over N^4}P-5v^{(4)}P \nonumber \\
&-&10v^{(3)}P^2-10v^{(2)}P^3-P^5.
\end{eqnarray}

The joint default probabilities are given by the integrals in equation (\ref{key}) which we evaluate by numerical integration. For this we need the inverse and determinant of a $m \times m$ correlation matrix ($m \le N$) with off diagonal elements $\xi$. It has one eigenvalue\footnote{For large portfolios this correlation matrix is only consistent for $\xi$ non-negative since a correlation matrix must have non-negative eigenvalues. It is very important that the correlation matrix $\xi_{ij}$ is positive semi-definite. If it has negative eigenvalues the integrals in equations (\ref{key}) are not well defined. Typically a correlation matrix that is forecast using qualitative methods will have some negative eigenvalues. A practical method for constructing the consistent correlation matrix that is closest to a forecasted one is given in Rebonato and  J\"{a}ckel (2000).
} equal to $1+(m-1)\xi$ and the others equal to $1-\xi$. Consequently its determinant is 
\be
{\rm det} [\xi_{ij}]=(1-\xi)^{m-1}[1+(m-1) \xi].
\ee
Its inverse has diagonal elements,
\be
\xi^{(-1)}_{ii}={1+(m-2) \xi \over 1+(m-2) \xi -(m-1)\xi^2},
\ee
and off diagonal elements ($i \ne j$),
\be
\xi^{(-1)}_{ij}=-{\xi \over 1+(m-2)\xi-(m-1)\xi^2}.
\ee
In Table V the joint default probabilities are given for $P=2\%$.
\begin{center}
Table V:  Joint Default Probabilities when $P=2\%$.
\vskip 0.25in
$
\bordermatrix{& P_{12}\times 10^3 & P_{123}\times 10^3& P_{1234}\times 10^3 & P_{12345}\times 10^3 &\xi  \cr
& 11.6 & 8.74 &7.25 & 6.32 &0.90   \cr 
 & 9.85 & 6.78 & 5.28 & 4.39 &0.85 \cr 
 & 8.45 & 5.31 & 3.89 & 3.09 &0.80  \cr
 & 7.28 & 4.17 & 2.86 & 2.16  &0.75 \cr
 & 6.28 & 3.27 & 2.09& 1.50 &0.70 \cr
 & 5.41 & 2.54 & 1.51 & 1.02 &0.65 \cr
 & 4.64 & 1.96 & 1.07& 0.679 &0.60 \cr
 & 3.98 & 1.49 & 0.746 & 0.439 &0.55 \cr 
 & 3.39 & 1.12 & 0.504 & 0.273 &0.50\cr
 & 2.87 & 0.820 & 0.330& 0.162 & 0.45 \cr
 & 2.41 & 0.589 & 0.208& 9.13 \times 10^{-2}& 0.40\cr
 & 2.01 & 0.412& 0.124 & 4.79 \times 10^{-2}& 0.35 \cr
 & 1.66 & 0.279 & 7.01\times 10^{-2}& 2.30\times 10^{-2} & 0.30\cr
 & 1.36 & 0.181 & 3.67\times 10^{-2} & 9.85 \times 10^{-3} &0.25\cr
}$
\end{center}
Using equations (\ref{mom2}) to (\ref{mom5}) and the results of Table V the moments of the default loss probability distribution can be calculated. They are given in Table VI for the case $N=100$. 
\begin{center}
Table VI:  Moments of the Default Loss Probability Distribution when $P=2\%$ and $N=100$.
\vskip 0.25in
$
\bordermatrix{& v^{(2)}\times 10^3 & v^{(3)}\times 10^3& v^{(4)}\times 10^3 & v^{(5)}\times 10^3 &\xi  \cr
& 11.3 & 8.14 &6.66 & 5.71 &0.90   \cr 
 & 9.55 & 6.29 & 4.85 & 3.97 &0.85 \cr 
 & 8.17 & 4.91 & 3.56 & 2.79 &0.80  \cr
 & 7.01 & 3.84 & 2.62 & 1.95 &0.75 \cr
 & 6.01 & 2.99 & 1.91& 1.35 &0.70 \cr
 & 5.15 & 2.31 & 1.38 & 0.923 &0.65 \cr
 & 4.40 & 1.77 & 0.975& 0.614 &0.60 \cr
 & 3.74 & 1.33 & 0.675 & 0.397 &0.55 \cr 
 & 3.15 & 0.988 & 0.455 & 0.247 &0.50\cr
 & 2.64 & 0.717 & 0.297& 0.147 & 0.45 \cr
 & 2.19 & 0.506 & 0.185& 8.27 \times 10^{-2}& 0.40\cr
 & 1.79 & 0.346& 0.110 & 4.34 \times 10^{-2}& 0.35 \cr
 & 1.45 & 0.227 & 6.16\times 10^{-2}& 2.09\times 10^{-2} & 0.30\cr
 & 1.15 & 0.142 & 3.18\times 10^{-2} & 8.97 \times 10^{-3} &0.25\cr
}$
\end{center}
For $\xi = 0.25$ the skewness of the default loss probability distribution is 3.6 indicating that for $\xi \ge 0.25$ this probability distribution is highly nontrivial and is very far from a normal distribution. Despite this the moment expansion is a useful tool for determining $\alpha_{opt}$.

\end{document}